\documentclass{article}

\usepackage{arxiv}

\usepackage[utf8]{inputenc} 
\usepackage[T1]{fontenc}    
\usepackage{url}            
\usepackage{booktabs}       
\usepackage{amsfonts}       
\usepackage{amsmath}        
\usepackage{nicefrac}       
\usepackage{microtype}      
\usepackage{algorithm}      
\usepackage{algorithmic}
\usepackage{graphicx}
\usepackage{natbib}
\usepackage{doi}

\usepackage[table]{xcolor}  
\usepackage{listings}       

\definecolor{codegreen}{rgb}{0,0.6,0}
\definecolor{codegray}{rgb}{0.5,0.5,0.5}
\definecolor{codepurple}{rgb}{0.58,0,0.82}
\definecolor{backcolour}{rgb}{0.95,0.95,0.92}

\lstdefinestyle{mystyle}{
    backgroundcolor=\color{backcolour},   
    commentstyle=\color{codegreen},
    keywordstyle=\color{magenta},
    numberstyle=\tiny\color{codegray},
    stringstyle=\color{codepurple},
    basicstyle=\ttfamily\footnotesize,
    breakatwhitespace=false,         
    breaklines=true,                 
    captionpos=b,                    
    keepspaces=true,                 
    numbers=left,                    
    numbersep=5pt,                  
    showspaces=false,                
    showstringspaces=false,
    showtabs=false,                  
    tabsize=2
}

\newcommand{\code}[1]{\textit{#1}}
\newcommand{\pkg}[1]{\textit{#1}}
\newcommand{\proglang}[1]{\textit{#1}}
\newcommand{\class}[1]{\textit{#1}}
\newcommand{\module}[1]{\textit{#1}}

\title{Rapid Experimentation with Python Considering Optional and Hierarchical Inputs}

\author{ \href{https://orcid.org/0000-0002-0063-1254}{\includegraphics[scale=0.06]{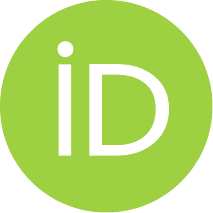}\hspace{1mm}Neil C.~Ranly} \\
	Department of Operational Sciences\\
	Air Force Institute of Technology\\
	Wright-Patterson AFB, OH 45433  \\
	\texttt{neil.ranly@afit.edu} \\
	\And
	\href{https://orcid.org/0000-0002-1039-1055}{\includegraphics[scale=0.06]{orcid.pdf}\hspace{1mm}Torrey D.~Wagner} \\
	Department of Systems Engineering and Management\\
	Air Force Institute of Technology\\
	Wright-Patterson AFB, OH 45433 \\
}

\renewcommand{\shorttitle}{\textit{raxpy} Rapid Experimentation with Python}

\hypersetup{
pdftitle={Rapid Experimentation with Python},
pdfauthor={Neil Ranly, Torrey Wagner},
pdfkeywords={space-filling design, computer experiment, maximum projection criterion, hierarchical experimentation factors},
}

\begin{document}
\maketitle

\begin{abstract}
Space-filling experimental design techniques are commonly used in many computer modeling and simulation studies to explore the effects of inputs on outputs. This research presents \pkg{raxpy}, a \proglang{Python} package that leverages expressive annotation of \proglang{Python} functions and classes to simplify space-filling experimentation. It incorporates code introspection to derive a \proglang{Python} function's input space and novel algorithms to automate the design of space-filling experiments for spaces with optional and hierarchical input dimensions. In this paper, we review the criteria for design evaluation given these types of dimensions and compare the proposed algorithms with numerical experiments. The results demonstrate the ability of the proposed algorithms to create improved space-filling experiment designs. The package includes support for parallelism and distributed execution. \pkg{raxpy} is available as free and open-source software under a MIT license.
\end{abstract}


\section{Introduction}

Design of experiments (DOE) is a branch of applied statistics that generates a set of values, or simply the design points, as inputs to a function. By passing each point through the function and assembling a dataset with the function-returned outputs, experimenters perform analysis to gain insights into how the inputs affect outputs. In many situations, given constrained resources and non-trivial execution durations of these functions, experimenters are constrained on the number of points they can consider. The objective in designing these experiments is to generate a set of points that maximizes the insight gained from the experiment's outputs and that meets the experimenter's objectives given this point-set-size constraint.

This research focuses on DOEs to support exploration-based objectives of computer-based executable functions that possess either non-linear response surfaces, multiple output variables, or system-of-system emergent dynamics. Computer-executable-functions, such as computer-based simulation models \citep{sanchez_work_2020}, are a growing source of data for deep learning \citep{makoviychuk_isaac_2021} and a proven method for generating insights into rare and unrealized situations. DOEs for exploration use cases include the exploration of a input-space at the start of an optimization effort, the discovery of non-linear relationships between inputs and outputs \citep{box_design_1959}, the validation of a complex simulation model with multiple outputs \citep{tolk_better_2017}, comparing different versions of computer-executable-functions in a black-box manner, and the generation of a dataset supporting downstream machine learning, such as surrogate modeling \citep{arboretti_design_2022,fontana_design_2023}.

Common DOE techniques such as grid sampling and random sampling have undesirable properties for exploration use cases. Researchers may find that only a small subset of input dimensions have an affect on the outputs, known as the effect sparsity principle \citep{young_graphical_2024}. In these cases, grid-based designs and many factorial-based design-variations possess projection properties that inefficiently duplicate the execution of points that collapse to a same lower-dimensional-point after dropping the input dimensions that lack practical-significance on the outputs \citep{joseph_space-filling_2016}. For example, consider the two-dimension, two-level full-factorial design of $\{(0,0), (0,1), (1,0), (1,1)\}$ projecting to the design $\{(0), (0), (1), (1)\}$ with duplicate points after removing the 2nd dimension. In addition, grid and factorial based design sizes increase exponentially as the number of dimensions increase making them infeasible to consider for large dimensional input spaces. Random sampling designs, by virtue of being stochastic, often fail to hold space-filling properties and may possess highly correlated design points given design size constraints. Correlations among the inputs confound the determination of the true cause of output differences. These undesirable theoretical properties of these designs can cause significant practical differences when compared to alternative design-generation algorithms that attempt to evenly-sample or uniformly-sample the design-space; this paper refers to these alternative algorithms as space-filling DOEs. Evaluations of space-filling DOEs have shown better performance in many surrogate modeling benchmarks \citep{arboretti_design_2022,fontana_design_2023}.

Initial research into space-filling designs, such as Latin-hypercube sampling based designs (LHDs), assumes bounded, continuous input dimensions, i.e., rectangular input spaces \citep{johnson_minimax_1990, owen_orthogonal_1992, damblin_numerical_2013}. LHDs ensure any subsequent dropping of dimensions from a input dataset retains all points uniqueness. LHDs ensure this by generating $n$ unique values for each input dimension; where $n$ is the desired point set size given resource limitations and function-execution budgets. LHDs considers the range of an input dimension and creates $n$ evenly, mutual exclusive sub-bounded regions and chooses a centered value or a random value within this sub-bounded region. In order to avoid undesirable correlations and distances between points, optimization algorithms derive the final input value combinations from these single dimension value sets \citep{damblin_numerical_2013, joseph_space-filling_2016}. 

Additional space-filling design research investigates non-rectangular input spaces, discrete-valued dimensions, and additional constraints on the input space. Maximum projection space-filling design techniques support discrete-level numeric inputs while maximizing the sub-space projection distance of points \citep{joseph_designing_2020, lekivetz_fast_2019}. Fast-flexible filling techniques \citep{lekivetz_fast_2015, chen_optimal_2019}, adaptive optimization techniques \citep{wu_space-filling_2019}, mixed integer non-linear programming techniques \citep{ozdemir_robust_2023}, and partial swarm optimization techniques \citep{chen_particle_2022} address non-linear constraints and non-convex volumes. Orthogonal uniform composite designs provide techniques for flexible design sizes while retaining orthogonal properties of the design \citep{zhang_orthogonal_2020}. Sliced Latin hypercube designs \citep{qian_sliced_2012}, marginally coupled designs \citep{deng_design_2015}, and maximum-projection techniques \citep{joseph_designing_2020} address qualitative input dimensions, such as a categorical input factors. Techniques for sequential space-filling DOEs remove the requirement that the user needs to specify the design size in a upfront manner, enabling the dynamic consideration of time constraints or other insight criteria discovered during experimentation \citep{crombecq_efficient_2011, sheikholeslami_progressive_2017, parker_sequentially_2024}. \citet{wang_constrained_2021} provide sequential space-filling design techniques while considering input spaces with non-linear constraints. \citet{lu_practical_2022} discusses research methods to prioritize regions of the input space. 

A focus of this research is algorithms to generate space-filling designs for input spaces with optional-and-hierarchical dimensions. For example, consider experimenting with a object-detection computer-simulation with the input-space being the scene context and the inclusion of different types of objects, each type of object having additional input dimensions that are desired to vary to explore their influence on the system's detection performance. In this scenario, the inclusion of each object type represents an optional dimension of a simulation run. In addition, if the object type is included in the scene, then attributes related to the object, such as color, movement, position, and other object attributes, can be varied as part of the DOE.

Input spaces with these types of dimensions also occur in hyper-parameter search spaces \citep{bischl_mlrmbo_2017}. The Tree-structured Parzen Estimator is a hyper-parameter optimization technique supporting hierarchical spaces, studied with an initial random exploration design \citep{NIPS2011_86e8f7ab}. Decomposition search strategies to conduct optimization over similar input spaces have also been proposed \citep{li_volcanoml_2023}. From our literature review, a gap exists of algorithms to support space-filling designs for input spaces with optional-and-hierarchical dimensions. This research proposes space-filling DOE algorithms for these types of input spaces given a user design point-size trial budget, $n$.

Space-filling DOE algorithms are provided in commercial software and open-source software. \pkg{scipy} \citep{2020SciPy-NMeth} and \pkg{UniDOE} provide algorithms that maximize designs' discrepancy \citep{zhou2013mixture}. \pkg{MaxPro} \citep{shan_ba_and_v_roshan_joseph_maxpro_2015} provides algorithms that maximize the multi-dimension sub-projections of designs. \pkg{SlicedLHD} provides an implementation of Sliced Latin Hypercube Designs \citep{kumar_slicedlhd_2024}. \citet{noauthor_jmp_1989} provides a variety of algorithms including Fast-Flexible filling \citep{lekivetz_fast_2019} to support designs with non-linear constraints. See \citet{lucas_variability_2023} for a comparison of the space-filling properties of designs created with popular experimentation support software. In addition to the algorithm gap perceived, this research also addresses the software gap for the implementation of these techniques on computer-executable function experimentation subjects.

This research presents \pkg{raxpy}, a \proglang{Python} library that contributes novel methods to express experiments' input spaces with \proglang{Python}-type specifications and novel extensions to DOE space-filling algorithms to generate experiment designs that address optional-and-hierarchical input dimensions. \pkg{raxpy}'s use of \proglang{Python} enables instrumented execution of many types of computer-based executable functions not natively expressed in \proglang{Python}, such as command-line-based simulation programs, generative artificial intelligence web services, and other types of external web services. The proposed library enables rapid execution of space-filling experiments with the capabilities of \proglang{Python} and its ecosystem of libraries such that generated data can be quickly processed to enable downstream analysis activities.

Section \ref{sec:design-space-specification} discusses how \pkg{raxpy} is used to perform experiments and create designs. Section \ref{sec:designing-experiments} discusses how \pkg{raxpy} evaluates and creates DOEs given optional-and-hierarchical dimensions. Finally, Section \ref{sec:analysis} discusses the results of numerical experiments comparing the space-filling properties for designs generated with the proposed algorithms.

\section{Performing Experiments} \label{sec:design-space-specification}

The following code demonstrates how users of \pkg{raxpy} can create and execute a space-filling experiment with $n = 10$ points on an annotated \proglang{Python} function, such as the example function \code{f5} listed in section \ref{sec:function-annotations}.

\begin{lstlisting}[language=Python]
import raxpy

inputs, outputs = raxpy.perform_experiment(f5, 10)
\end{lstlisting}

The method \code{perform\_experiment} executes the following steps. First, it derives the input space from the annotated parameters of the passed-in function; function annotations are described subsequently in section \ref{sec:function-annotations}. The function's input space is represented as a \class{Space} object composed of zero-or-many \class{Dimension} objects; the \module{raxpy.spaces} module is described in section \ref{sec:design-spaces}. Secondly, the \class{Space} object for a function's input space is passed to a space-filling DOE algorithm that creates a DOE; the DOE algorithm is described in section \ref{sec:design-algorithms}. Next, the design points are mapped to the function parameters and executed for each point. Finally, it returns the points and the corresponding outputs. This automated mapping from a function signature, to input-space, to DOE, to function arguments for execution, and finally execution of function with arguments can increase the efficiency of conducting exploration experiments compared to manually mapping the results between each step.

\pkg{raxpy} can also create an experiment design without executing it, given an annotated function:

\begin{lstlisting}[language=Python]
doe = raxpy.design_experiment(f5, 10)
\end{lstlisting}

\subsection{Function Annotation} \label{sec:function-annotations}

\pkg{raxpy} utilizes the type specifications and annotations of \proglang{Python} functions and classes to express and to derive an experiment design's input and output space. Type specifications in programming languages have a long tradition of supporting expressive expectations of variables' types. Programmers have long valued type specification in large code bases to clearly express interfaces' input and output specifications to themselves and other programmers. Code compiling and static-code-analysis linting techniques also use type specifications to identify interface violations before software is executed. In many scripting languages, including \proglang{Python} 2, type specification is not required and is not supported. As the use of \proglang{Python} grew, community efforts and demands to enable code quality processes resulted in type specifications being incorporated to \proglang{Python} in version 3.5 as part of Python Enhancement Proposal (PEP) 484 \citep{python_pep484}. Since, many \proglang{Python} libraries have incorporated type specifications supplying common documentation of variable type expectations and enabling automated pre-runtime code quality processes. The following code demonstrates a function with three parameters with and without type specifications.

\begin{lstlisting}[language=Python]
from typing import Optional

# simple function without type hints
def f1(x1, x2, x3):
    return x1 * x2 if x3 is None else x1 * x2 * x3

# simple function with type hints
def f2(x1: float, x2:int, x3:Optional[float]) -> float:
    return x1 * x2 if x3 is None else x1 * x2 * x3
\end{lstlisting}

With the implementation of PEP 593 in \proglang{Python} 3.9, \proglang{Python} extends the concept of type specification to permit the annotation with additional meta-data \citep{python_pep593}. The follow code demonstrates a \proglang{Python} function's parameter type specifications with annotated meta-data; the first two parameters have a lower and upper bound specified, while the \code{x3} parameter has a value set specified to indicate the discrete-numeric values for this parameter.

\begin{lstlisting}[language=Python]
import raxpy
from typing import Annotated

# simple function with type hints and annotations
def f3(
    x1: Annotated[float, raxpy.Float(lb=0.0, ub=10.0)],
    x2: Annotated[float, raxpy.Float(lb=0.0, ub=2.0)],
    x3: Annotated[Optional[float], raxpy.Float(value_set=[0.0, 1.5, 3.0])]
) -> Annotated[float, raxpy.Float(tags=[raxpy.tags.MAXIMIZE])]:
    return x1 * x2 if x3 is None else x1 * x2 * x3 
\end{lstlisting}

In addition to the improved documentation, there are additional benefits for annotating variables. They also enable runtime validation of values beyond just the validation of a values' types; see the code below how \pkg{raxpy} can enable this. The main benefit, which we utilize in \pkg{raxpy}, is the introspection of a function signature to dynamically extract functions' design spaces. With the derived function's input and output space specification, \pkg{raxpy} analyzes the space's dimensions to design experiments, execute designs, and gather the results. Examples also demonstrate how this introspection of \pkg{raxpy} annotations can be adapted for use with other \proglang{Python} modules to perform optimization experiments. The expression of the experimentation subject as a \proglang{Python} function also enables software engineering inspired test-driven-development best practices such as unit-testing.

\begin{lstlisting}[language=Python]
# simple function with annotations and runtime validation
@raxpy.validate_at_runtime(check_outputs=False)
def f4(
    x1: Annotated[float, raxpy.Float(lb=0.0, ub=10.0)],
    x2: Annotated[int, raxpy.Float(lb=0.0, ub=2.0)],
    x3: Annotated[Optional[float], raxpy.Float(value_set=[0.0, 1.5, 3.0])]
) -> float:
    return x1 * x2 if x3 is None else x1 * x2 * x3

f4(3.14, 1, None) # no error
f4(3.14, 11, None) # runtime error given 11 value does not fall within range
\end{lstlisting}

To specify the input space of a \proglang{Python} function, each parameter of a function is annotated. \pkg{raxpy} code introspection algorithm checks for standard type specifications in the \pkg{typing} package to indicate \code{Optional} and \code{Union} parameters. This encourages experimenters to reuse existing type-specification best practices while minimizing the input space specification in an additional configuration section of the code or an external file. It also minimizes the experimenter's requirement to employ a different space-specification method. \pkg{raxpy} uses \pkg{dataclasses} to support the expression of hierarchical dimensions. The following code demonstrates how \pkg{raxpy} creates and executes a space-filling experiment with 25 points on an annotated function \proglang{Python} with optional hierarchical dimensions expressed with \code{typing.Union}, \code{typing.Optional}, and \code{dataclasses}.

\begin{lstlisting}[language=Python]
from dataclasses import dataclass

@dataclass
class HierarchicalFactorOne:
    x4: Annotated[float, raxpy.Float(lb=0.0, ub=1.0)]
    x5: Annotated[float, raxpy.Float(lb=0.0, ub=2.0)]

@dataclass
class HierarchicalFactorTwo:
    x6: Annotated[float, raxpy.Float(lb=0.0, ub=1.0)]
    x7: Annotated[float, raxpy.Float(lb=-1.0, ub=1.0)]


def f5(
    x1: Annotated[float, raxpy.Float(lb=0.0, ub=1.0)],
    x2: Annotated[Optional[float], raxpy.Float(lb=-1.0, ub=1.0)],
    x3: Union[HierarchicalFactorOne, HierarchicalFactorTwo],
):
    # placeholder for f5 logic
    return 1 
inputs, outputs = raxpy.perform_experiment(f5, 25)
\end{lstlisting}

The annotation code for the \pkg{raxpy} type specification is organized in the \module{raxpy.annotations} module. Although we expect many uses of \pkg{raxpy} to employ the \module{raxpy.annotations} module or existing annotation libraries, the \module{raxpy.spaces} module provides common data structures for DOE algorithms and external specifications of a space. For example, a user interface may permit an experimenter to specify the input dimensions of a space for a \proglang{Python} function that dynamically configures a simulation model given the input provided. 

\subsection{Design Spaces} \label{sec:design-spaces}

A \proglang{Python} function corresponds to two design spaces, an input space derived from the input parameters of the function, and an output space derived from the return type specification. The code below shows the ability to specify a \class{InputSpace} without deriving it from a function's signature. A major difference between \pkg{raxpy} and the \pkg{ConfigSpace} \proglang{Python} package \citep{lind2019}, is the incorporation of an nullable attribute for dimensions. Dimension's nullable attributes enables a clear mapping from \proglang{Python} functions' parameters specified as optional. 

\begin{lstlisting}[language=Python]
space = raxpy.spaces.InputSpace(
    dimensions=[
        raxpy.spaces.Float(id="x1", lb=0.0, ub=1.0, portion_null=0.0),
        raxpy.spaces.Float(id="x2", lb=0.0, ub=1.0, portion_null=0.0),
        raxpy.spaces.Float(id="x3", lb=0.0, ub=1.0, nullable=True, portion_null=0.1),
    ]
)
\end{lstlisting}

The following dimension-types are also provided to support the direct mapping from complex function parameters:
\begin{itemize}
    \item \code{Composite} objects represent dimensions mapped from \code{dataclass} type specification;
    \item \code{Variant} objects represent dimensions mapped from \pkg{typing} \code{Union} type specification;
    \item \code{ListDim} objects represent dimensions formed from type specifications of from \pkg{typing} \code{List} type specification.
\end{itemize}

\section{Designing Experiments with Optional and Hierarchical Inputs} \label{sec:designing-experiments}

A common approach to creating space-filling designs is to measure and optimize a design with respect to a space-filling metric. Optional and hierarchical dimensions complicate the direct application of these methods. Section \ref{sec:criteria} suggests extensions to past space-filling measurement criteria and concepts to address these more complex space attributes. \pkg{raxpy} code examples are also provided demonstrating techniques to measure and analyze space-filling designs. Section \ref{sec:design-algorithms} presents DOE algorithms employing these concepts.

\subsection{Space Filling Metrics} \label{sec:criteria}

Seven space-filling metrics are extended and used in this work and described in this section.

\begin{itemize}
    \item $M^\text{ocov}$: Full-sub-space optionality-coverage percentage
    \item $M^\text{idis}$: Minimum interpoint distance
    \item $M^\text{adis}$: Average minimum single-dimension projection distance
    \item $M^\text{wdsr}$: Weight and sum of full-sub-space discrepancies
    \item $M^\text{sdsr}$: Star discrepancies, incorporating a null-region within dimensions
    \item $M^\text{maxpro}$: Variation of MaxPro metric extended to support optional and hierarchical dimensions
    \item DOE target allocation difference: Count of point allocations diverging from the target full-sub-space allocation counts
\end{itemize}

Let $X$ denote a DOE with $n$ $d$-dimensional points and $S$ denote the input-space such that $X \subseteq S$ and $S = (D_1, D_2, \dots, D_d)$ where $D_k$ denotes the acceptable range-of-values for dimension $k$. Let $D = \{1,2, \ldots, d\}$ represent the set of indices corresponding to the dimensions. Let $D^{\text{optional}} \subseteq D$ represent the set of indices corresponding to the optional dimensions in $S$ such that $\{\text{null}\} \subseteq D_k, \forall k \in D^{\text{optional}}$. Let $D^{\text{parent}} \subseteq D$ indicate the set of indices for hierarchical activation dimensions and $\{0,1\} \subseteq D_k, \forall k \in D^{\text{parent}}$. Let $D^{\text{real}} \subseteq D$ denote the set of indices corresponding to bounded real numbers, such that $[l_k, u_k] \subseteq D_k, \forall k \in D^{\text{real}}$. All numeric experiments and subsequent notation fix $l_k = 0$ and $u_k = 1$. Let $x_i$ represent $i$th point of $X$ and $x_{ik}$ represent point $i$'s value for dimension $k$. Let $P \subseteq D^{\text{parent}} \times D \times D_p $ denote a set of hierarchical constraints such that if $ (p,c,v) \in P$ then the child dimension $c$ for a point is constrained to be $\text{null}$ if the parent dimension $p$'s value is not $v$. This implies $c \in D^{\text{optional}}, \forall (p,c,v) \in P$.

The first criteria we consider is the extent of a design's coverage over the different possible combinations of optional parameters. To compute, we define a projection of $X$ to a $d$-dimensional binary space indicating the non-null specification of values in points as the optional-definition design projection, $X^\text{opt}$. $X^\text{opt} \subseteq \{0, 1\}^{n \times d} $, where each element $ x_{ij}^\text{opt} $ is defined as:

\begin{equation}
x_{ij}^\text{opt} = \begin{cases}
0, & \text{if } x_{ij} = \text{null} \\
1, & \text{if } x_{ij} \neq \text{null}
\end{cases}
\end{equation}

Figure \ref{fig:full-sub-designs} visualizes this mapping for a example DOE of five dimensions where $D^{\text{optional}} = \{2,3,4,5\}, D^{\text{parent}} = \{3\}, D^{\text{real}} = \{1,2,4,5\}, P = \{(3,4,1),(3,5,1)\} $. We can compute the optionality-coverage percentage by taking the set-size of unique points $|\bigcup_{i=1}^{n} \{ x_i^\text{opt} \}|$ over the size of power set of $D^{\text{optional}}$, $|\mathcal{P}_P(D^{\text{optional}})|$. $\mathcal{P}$ is a slightly modified version of the power-set since some optional dimensions are only feasible given the hierarchical constraints specified with $P$ and some dimensions are constrained to be not-null, thus are active in every sub-space. \pkg{raxpy} uses a tree traversal technique to compute $\mathcal{P}_P(D^{\text{optional}})$. For the example depicted in Figure \ref{fig:full-sub-designs}, $\mathcal{P}_P(D^{\text{optional}}) = \{(1), (1,2), (1,2,3), (1,2,3,4), (1,2,3,5), (1,2,3,4,5), (1,3), (1,3,4), (1,3,5), (1,3,4,5)\}$. The depicted DOE in Figure \ref{fig:full-sub-designs} only samples six of these possible sub-spaces. We define full-sub-spaces (FSSs) as the set of sub-spaces indicated by $\mathcal{P}_P(D^{\text{optional}})$. Each point of $S$ maps to a single element of the $\mathcal{P}_P(D^{\text{optional}})$ set. For example points (rows) two and three in Figure \ref{fig:full-sub-designs} map to the full-sub-space of $\{1,2\}$.

Given a \code{DesignOfExperiment} object named \code{doe}, \pkg{raxpy} can compute optionality-coverage percentage:

\begin{lstlisting}[language=Python]
opt_coverage = raxpy.measure.compute_opt_coverage(doe)
\end{lstlisting}

Given a \code{InputSpace} object named \code{space}, \pkg{raxpy} can compute $\mathcal{P}_P(D^{\text{optional}})$:

\begin{lstlisting}[language=Python]
fss_dim_sets = space.derive_full_subspaces()
\end{lstlisting}

\begin{figure}[t!]
\centering
\includegraphics[width=\linewidth]{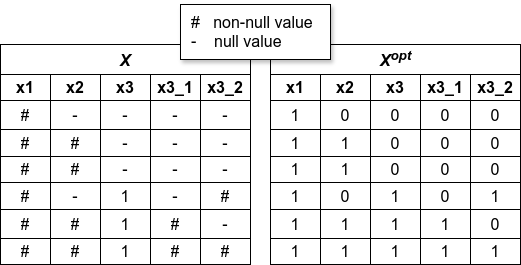}
\caption{\label{fig:full-sub-designs} $X$ to $X^\text{opt}$ point mapping}
\end{figure}

The second criteria we employ is a design's minimum interpoint distance, $M^\text{idis}$ \citep{johnson_minimax_1990}.

\begin{equation}
\text{min}_{x_i, x_j \in X} (\sum_k^d{\text{dist}_k (x_{i} - x_{j})^p})^{(1/p)}
\end{equation}

This criteria is motivated by the objective to maximize the minimal distance between all the points. To support measurement of designs with optional dimensions, we define the distance computations with a revised distance equation for a 0-1 normalized, encoded design:

\begin{equation}
   \text{dist}_{k}(x_i, x_j) =
   \begin{cases}
   |x_{ik} - x_{ik}| & \text{if } x_{ik} \neq \text{null} \text{ and } x_{jk} \neq \text{null}, \\
   1 & \text{if } (x_{ik} = \text{null} \text{ and } x_{jk} \neq \text{null}) \text{ or } (x_{jk} = \text{null} \text{ and } x_{ik} \neq \text{null}), \\
   0 & \text{if } x_{ik} = \text{null} \text{ and } x_{jk} = \text{null}.
   \end{cases}
\end{equation}

Given a \class{DesignOfExperiment} object named \code{doe}, \pkg{raxpy} provides support for this:

\begin{lstlisting}[language=Python]
min_interpoint_dist = raxpy.measure.compute_min_interpoint_dist(doe, p=2)
\end{lstlisting}

The discrepancy criteria evaluates the extent a design achieves uniform distribution across a bounded, numeric-value input space \citep{zhou2013mixture}. We consider two discrepancy measurement extensions. The first extension $M^\text{wdsr}$, we measure the design by decomposing the design to well-bounded, non-null sub-input-spaces. We weight and sum the discrepancies of the sub-designs, with weights representing the target number of points allocated to each FSS over $n$. We consider this metric especially applicable for design evaluation when subsequent experimentation is expected to focus on one sub-space given the initial exploration results. We only employ this metric for design comparisons given designs with the same FSS point allocations. 

The second discrepancy extension, $M^\text{sdsr}$, we consider is the direct incorporation of $\text{null}$ regions in $D^{\text{optional}}$ dimensions as part of the uniform distribution computation considered by discrepancy. Discrepancy is defined as the maximum difference between the portion of design points within a region and the volume of that same region as a ratio to the volume of the whole space. To compute with nullable dimensions, we consider null values to directly precede 0 on the number line and the null region sizes of a dimension to be specified as stated in section \ref{sec:design_heuristics} or specified by the user.

\begin{equation}
    \sup_{u \in [\text{null},1]^d} \Big| \frac{|\{x : x \leq u\}|}{n} - \prod_{k=1}^d \Big(\bar{\alpha_k} + u_k (1 - \bar{\alpha_k} )\Big) \Big|
\end{equation}

where $\bar{\alpha_k}$ indicates the size of the null region for dimension $k$. Future research is suggested to explore mixture discrepancy \citep{zhou2013mixture} variation extensions.

\begin{lstlisting}[language=Python]
discrepancy = raxpy.measure.compute_star_discrepancy(doe)
\end{lstlisting}

To measure the single-dimensional projection properties of designs, we compute the average minimum distance, $M^\text{adis}$, between single-dimension projections, where $I_k^{\text{not-null}} \subseteq \{i : x_{ik} \neq \text{null}\}$ representing the indices of non-null values:

\begin{equation}
\frac{1}{d} \sum_{k=1}^d {\min_{i, j \in I_k^{\text{not-null}}, i \neq j}{|x_{ik}-x_{jk}|}}
\end{equation}

\begin{lstlisting}[language=Python]
avg_min_proj_dist = raxpy.measure.compute_average_dim_dist(doe, p=2)
\end{lstlisting}

To assess the multi-dimensional projection of designs, we propose a variation of the MaxPro metric. If $k \in D^{\text{optional}} \cup D^{\text{parent}}$, $\bar{\alpha}_k$ denotes $ 1 / \alpha_k$, where $\alpha_k$ denotes either $|D_k| + 1$ if $D_k$ is finite and is optional or an estimation of the dimension's complexity; see section \ref{sec:design_heuristics}. This avoids dividing by zero for designs that must contain duplicate values for dimensions with a finite set of values. Another effect of this is that differences corresponding to dimensions with less complexity are weighted less. Otherwise, $\bar{\alpha}_k = 0$ for bounded, non-optional real values to avoid duplicates. 

\begin{equation}
\left(\frac{1}{\binom{n}{2}} \sum_{i=1}^{n-1} \sum_{j=i+1}^{n} \frac{1}{\prod_{k=1}^d ({\text{dist}_k (x_{ik}, x_{jk}) + \bar{\alpha}_k )^2}}\right)^\frac{1}{d}
\end{equation}

\begin{lstlisting}[language=Python]
max_pro = raxpy.measure.compute_max_pro(doe)
\end{lstlisting}

\subsection{Design Algorithms} \label{sec:design-algorithms}

To support the construction of designs, we propose five algorithms and evaluate those algorithms in addition to a baseline random-design algorithm. Some algorithms leverage a traditional space-filling design algorithm (TSFD) for a real, 0-1 bounded space. The TSFD we employ for numerical experiments in section \ref{sec:analysis} is the \pkg{scipy.stats.qmc} \class{LatinHypercube} algorithm based on centered discrepancy optimization. Note that each proposed algorithms' design can also be post-optimized further with a MaxPro-based optimization algorithm, explained at the end of this section. This results in the comparison of 12 algorithms in section \ref{sec:analysis}.

\begin{enumerate}
    \item The Full-Subspace (FSS-LHD) algorithm mimics a manual approach of separately considering the individual full-sub-spaces and creating a separate space-filling design for each. Each FSS is allocated a portion of the $n$ points in the design. In the absence of a user-specified allocation, heuristics are used to determine the allocation given null-portion attributes; see section \ref{sec:design_heuristics}. \pkg{raxpy} also supports a feature to force at least one point allocated to each FSS if possible. This results in the creation of a full-factorial inspired $X^\text{opt}$ design, at least one point for each element of $\mathcal{P}_P(D^{\text{optional}})$. Next, a TSFD is used to generate designs for each full-sub-design. 

    \item A variation of the FSS-LHD algorithm above using random designs instead of space-filling designs (FSS-Random).

    \item The Full-Subspace-with-Value-Pool (FSS-LHD-VP) algorithm revises the FSS-LHD algorithm by computing the number of values needed for each dimension given target FSS point allocations. Given this point count for each dimension, it next creates a value-pool for the dimension using a LHD method to generate spaced values. Next, starting with the largest FSS with the largest point allocation, the appropriate number of values are pulled from the value-pool across the quantiles of the remaining values in the pool. The pull values are shuffled to create an initial sub-design and then passed to a space-filling centered discrepancy optimization algorithm based on the \pkg{scipy.stats.qmc} \class{LatinHypercube} algorithm.

    \item The Tree-Traversal-Design (TT-LHD) algorithm starts by creating a root design using a TSFD for the dimensions without any parents. The root 0-1 encoded design values are then projected to [0-1]+null space using the values under the null-portion threshold to project to the null value and values greater being rescaled to 0-1 values given the range from the null-portion threshold to 1. For $D^{\text{parent}}$ and finite numeric-value-set based dimensions, these values are mapped to values representing the discrete values in the finite set, in a similar manner as suggested in \citet{joseph_designing_2020}.  Next, the algorithm creates sub-designs with TSFD for children-dimension sets given common parents, repeating the same 0-1 mapping logic. The lower-level sub-designs are merged with the root-design using a centered discrepancy optimization algorithm, only revising the children-dimensional values to avoid undoing previous merge optimizations. It repeats this in a recursive manner into the deeper nodes of the hierarchical space.

    \item The Whole-Projection-Design (P-LHD) algorithm employs a TSFD to create an initial design for the flattened hierarchical space, initially ignoring the optional-and-hierarchical-properties of dimensions. Given this initial design, using a similar value-mapping logic as the TT-LHD algorithm above, the values for optional dimensions are projected to a [0-1]+null space as needed. Values projected to null for parent dimensions are horizontally extended onto all children dimensions.
\end{enumerate}

As a baseline to compare the performance of these algorithms, we also consider a completely random design algorithm that addresses null values for optional dimensions according to the null-portion attribute assigned to optional dimensions. This is labeled as "Random" in section \ref{sec:analysis}.

A design Simulated Annealing MaxPro optimization is provided and outlined below where $I_k^{\text{not-null}}$ indicates the indices of rows with non-null values for dimension $k$. This algorithm is designed to retain the initial design's $X^\text{opt}$ structure, retaining the user's or the design heuristic's FSS allocations. It fulfills this by only swapping rows' values in $I_k^{\text{not-null}}$. An algorithm using MaxPro optimization is labeled with the suffix "-MP" in section \ref{sec:analysis}.

\begin{algorithm}
\caption{Simulated Annealing MaxPro Design Optimization}
\label{alg:maxpro_opt}
\begin{algorithmic}[1]
\REQUIRE An initial design $X$
\REQUIRE Max-iterations $M$
\REQUIRE MaxPro Measurement Function $c_\text{MaxPro}$
\STATE Initialize $X^\text{best} \leftarrow X$
\FOR{$m = 1$ to $M$ }
    \STATE $t \leftarrow 1.0 - m/M$
    \STATE $k \sim \text{Uniform}(D^{\text{real}}) $
    \STATE $i ,j \sim \text{Uniform}(I_k^{\text{not-null}}), \text{Uniform}(I_k^{\text{not-null}})$
    \WHILE{ $i = j$ }
        \STATE $j \sim \text{Uniform}(I_k^{\text{not-null}})$
    \ENDWHILE
    \STATE $X^\text{try} \leftarrow \text{ColumnSwap}(X,i,j,k) $
    \STATE $P \leftarrow e^{-(c_\text{MaxPro}(X^\text{try})-c_\text{MaxPro}(X^\text{best}))/t}$
    \IF{ $ P < \text{Uniform}([0 - 1]) $}
        \STATE $X^\text{best} \leftarrow X^\text{try} $
    \ENDIF
\ENDFOR
\RETURN $X^\text{best}$
\end{algorithmic}
\end{algorithm}

\begin{figure}
        \centering
        \includegraphics[width=1\linewidth]{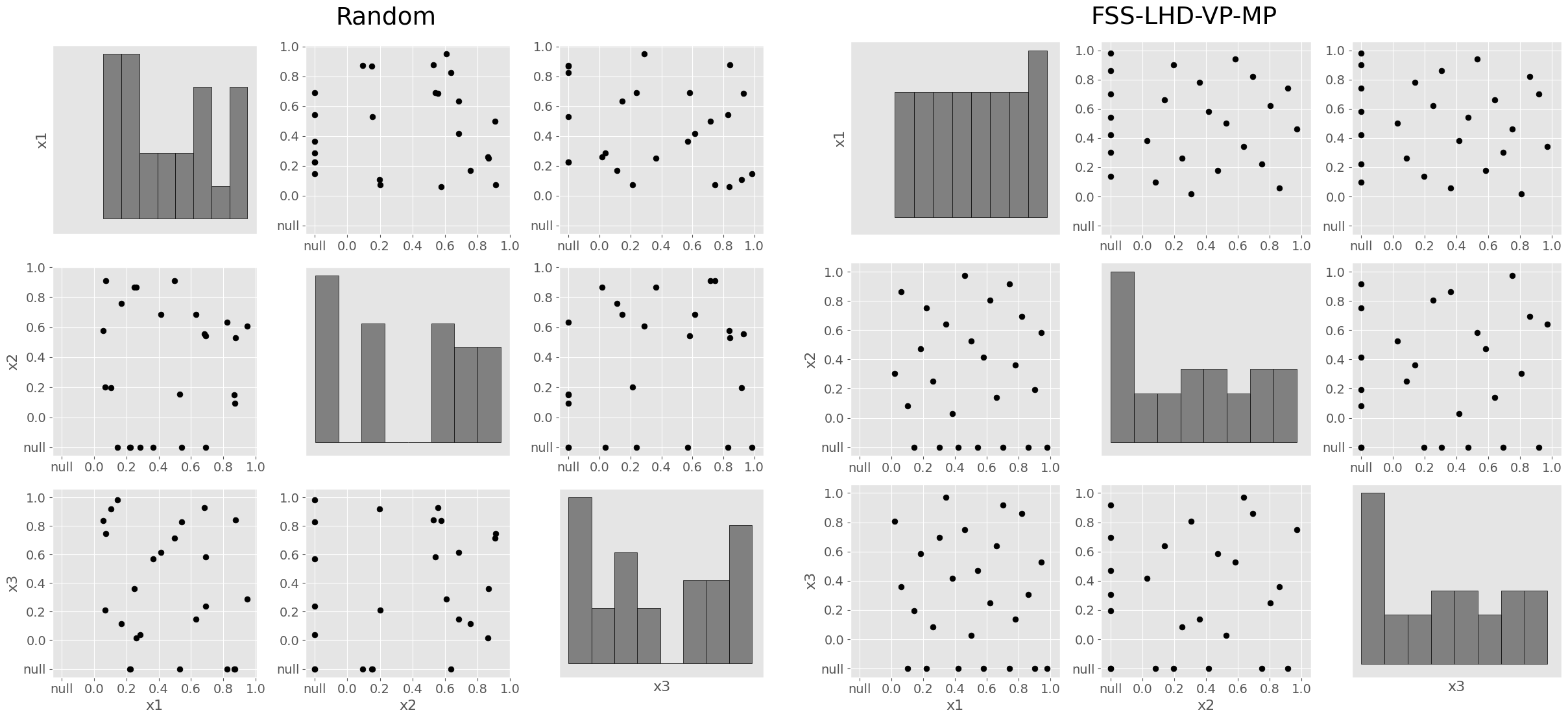}
        \caption{\label{fig:scatter-hist} Scatter-plots of the points from a random DOE (left) and a FSS-LHD-VP-MP DOE (right)}
\end{figure}

Figure \ref{fig:scatter-hist} presents the three-dimensional distribution of a 24-point experiment design space with a series of two-dimensional scatter plots and histograms. A random generated design is shown on the left, and a design generated with the FSS-LHD-VP-MP algorithm is shown on the right. The three dimensions are x1, x2 and x3, and were defined with a range of [0,1]; dimensions x2 and x3 are defined as optional. Histograms for each of the variables is shown along the diagonal of each plot. The ideal histogram is uniform for a space-filling design, and it is visible that the FSS-LHD-VP-MP histograms are far more uniform with a increased number of null values given the null region size.

\subsection{Design Heuristics} \label{sec:design_heuristics}

Each $D^{\text{optional}}$ dimension has an \code{null\_portion} attribute. This attribute indicates the size, from a input-space uniform distribution perspective, of the $\text{null}$ region in the dimension. This directly influences the percentage of points within a DOE that should have a $\text{null}$ value for this dimension if a user does not explicitly specify FSS allocations. \pkg{raxpy} utilizes heuristics to specify this attribute if not specified by the user. First, the number of levels of a dimension is estimated. We denote this simply as the dimension's complexity, $\alpha_k$. If a dimension is a finite set of values, the complexity is the set size. If a dimension $k \in D^{\text{real}}$, we currently specify the complexity to three plus one representing the $\text{null}$ value; one unit of the complexity represents the $\text{null}$ region. We then set \code{null\_portion}, $\bar{\alpha}_k$, to $1/\alpha_k$.

For the algorithms that utilize FSS point-allocation targets, we compute the percentage of points to allocate to the FSSs as needed. Let $S_{D^{\text{o}}}$ represent the FSS of $S$ given the dimensions specified in $D^{\text{o}} \in \mathcal{P}_P(D^{\text{optional}})$. Next, the percentage of points to allocate to each $S_{D^{\text{o}}}$ is computed as:

\begin{equation}
    \prod_{k \in D^\text{o}} (1 - \bar{\alpha_k})
\end{equation}

For design algorithms that utilize the full-sub-space allocation techniques, \pkg{raxpy} also allows user to directly provide their own FSS allocations.

\section{Analysis of Space-Filling Capability} \label{sec:analysis}

\subsection{Methodology}
We demonstrate the capabilities of \pkg{raxpy} with numeric experiments that compare designs' space-filling properties generated from the proposed algorithms. Each algorithm is considered by itself and also with the MaxPro optimization for a total of twelve design algorithms. We investigate four different input spaces (labeled 
"basic", "simple", "modest", and "complex") and three design sizes for each input space, and then generate 30 designs, i.e., replications, with each algorithm to analyze the variability of design measures. This results in 360 designs evaluated. 

Every design is evaluated with respect to $M^\text{idis}$, $M^\text{adis}$, $M^\text{sdcr}$, and $M^\text{wdcr}$. The algorithm replication evaluation means are ranked within each input space and design size combination. Full-sub-space allocation differences from target allocations are also computed and summarized.

The first input space, $S^\text{basic}$, we consider consists of $ D^{\text{real}} = \{1,2,3\}$, and $ D^{\text{optional}} = \{2,3\}$. For this input space, we consider design sizes $ n \in \{4|S^{\text{basic}}|, 8|S^{\text{basic}}|, 12|S^{\text{basic}}|\} = \{ 12, 24, 36 \}$. The input space is derived from the following function signature:

\begin{lstlisting}[language=Python]
def f_basic(
    x1: Annotated[float, raxpy.Float(lb=0.0, ub=1.0)],
    x2: Annotated[Optional[float], raxpy.Float(lb=0.0, ub=1.0)],
    x3: Annotated[Optional[float], raxpy.Float(lb=0.0, ub=1.0)]
):
\end{lstlisting}

The second input space, $S^\text{simple}$, we consider consists of $ D^{\text{real}} = \{1,2,4,5\}$, $ D^{\text{parent}} = \{3\}$, and $ D^{\text{optional}} = \{3,4,5\}$. For this input space, we consider $ n \in \{20, 40, 60\}$. The input space is derived from the following function signature:

\begin{lstlisting}[language=Python]
@dataclass
class HierarchicalFactorOne:
    x4: Annotated[float, raxpy.Float(lb=0.0, ub=1.0)]
    x5: Annotated[Optional[float], raxpy.Float(lb=0.0, ub=1.0)]

def f_simple(
    x1: Annotated[float, raxpy.Float(lb=0.0, ub=1.0)],
    x2: Annotated[Optional[float], raxpy.Float(lb=0.0, ub=1.0)],
    x3: Optional[HierarchicalFactorOne]
):
\end{lstlisting}

The third input space, $S^\text{modest}$, we consider consists of $ D^{\text{real}} = \{1,2,4,5,7,8\}$, $ D^{\text{parent}} = \{3,6\}$, and $ D^{\text{optional}} = \{2,3,4,5,6,7,8\}$. For this input space, we consider $ n \in \{32, 64, 96\}$. The input space is derived from the following function signature:

\begin{lstlisting}[language=Python]
@dataclass
class HierarchicalFactorTwo:
    x7: Annotated[float, raxpy.Float(lb=0.0, ub=1.0)]
    x8: Annotated[Optional[float], raxpy.Float(lb=0.0, ub=1.0)]

def f_modest(
    x1: Annotated[float, raxpy.Float(lb=0.0, ub=1.0)],
    x2: Annotated[Optional[float], raxpy.Float(lb=0.0, ub=1.0)],
    x3: Optional[HierarchicalFactorOne],
    x6: Optional[HierarchicalFactorTwo],
):
\end{lstlisting}

The fourth input space, $S^\text{complex}$, we consider consists of $ D^{\text{real}} = \{1,2,4,5,7,8,9,10\}$, $ D^{\text{parent}} = \{3,5, 8\}$, and $ D^{\text{optional}} = \{2,3,4,5,6,7,8,9,10\}$. For this input space, we consider $ n \in \{32, 64, 96\}$. The input space is derived from the following function signature:

\begin{lstlisting}[language=Python]
@dataclass
class HierarchicalFactorLevel2:
    x6: Annotated[float, raxpy.Float(lb=0.0, ub=1.0)]
    x7: Annotated[Optional[float], raxpy.Float(lb=0.0, ub=1.0)]

@dataclass
class HierarchicalFactorLevel1A:
    x4: Annotated[float, raxpy.Float(lb=0.0, ub=1.0)]
    x5: Optional[HierarchicalFactorLevel2]

@dataclass
class HierarchicalFactorLevel1B:
    x9: Annotated[float, raxpy.Float(lb=0.0, ub=1.0)]
    x10: Annotated[Optional[float], raxpy.Float(lb=0.0, ub=1.0)]

def f_complex(
    x1: Annotated[float, raxpy.Float(lb=0.0, ub=1.0)],
    x2: Annotated[Optional[float], raxpy.Float(lb=0.0, ub=1.0)],
    x3: Optional[HierarchicalFactorLevel1A],
    x8: Optional[HierarchicalFactorLevel1B],
):
\end{lstlisting}

\subsection{Results and Discussion}

Table \ref{tab:results} shows the resulting rankings of the algorithms by taking the average rankings across input space and design size combinations; ranking each algorithm on a 1-12 scale where 1 is the best. The top ranked algorithms are highlighted, with the best in the darkest shade of green. The FSS-LHD-VP-MP algorithm is in the top three performing algorithms over all four metrics and the best for $M^\text{idis}$. The additional processing with MaxPro optimization, designated with a -MP suffix, shows a trend with helping to improve the space-filling properties of the designs with respect to $M^\text{idis}$ and $M^\text{sdcr}$, while hurting $M^\text{wdcr}$ for the FSS-LHD and FSS-LHD-VP algorithms that originally optimized discrepancies by full-sub-spaces. 

\renewcommand{\arraystretch}{1.2} 
\begin{table}[H]
    \centering
    \begin{tabular}{|lcccc|}
         \hline
         Algorithm & $M^\text{idis}$ & $M^\text{adis}$ & $M^\text{sdcr}$ & $M^\text{wdcr}$ \\
         \hline
         Random* & 11.75 & 9.0 & 11.0 & 9.83 \\
         FSS-Random & 11.08 & 10.17 & 9.83 & 11.33 \\
         FSS-LHD & 5.67 & 7.83 & 7.42 & \cellcolor[HTML]{50c878} 1.58 \\
         FSS-LHD-VP & 5.5 & \cellcolor[HTML]{77dd77} 2.5 & 5.08 & \cellcolor[HTML]{77dd77} 2.67 \\
         TT-LHD* & 8.83 & \cellcolor[HTML]{50c878} 1.67 & 7.08 & 5.0 \\
         P-LHD* & 9.83 & 4.33 & 7.17 & 5.17 \\
         Random-MP* & 7.75 & 9.0 & 8.25 & 9.67 \\
         FSS-Random-MP & \cellcolor[HTML]{d0f0c0} 3.17 & 10.17 & 7.25 & 8.58 \\
         FSS-LHD-MP & \cellcolor[HTML]{77dd77} 2.33 & 7.83 & 5.33 & 6.0 \\
         FSS-LHD-VP-MP & \cellcolor[HTML]{50c878} 1.0 & \cellcolor[HTML]{77dd77} 2.5 &\cellcolor[HTML]{77dd77}  2.5 & \cellcolor[HTML]{d0f0c0} 4.25 \\
         TT-LHD-MP* & 5.58 & \cellcolor[HTML]{50c878} 1.67 & \cellcolor[HTML]{50c878} 2.25 & 7.17 \\
         P-LHD-MP* & 5.5 & 4.33 & \cellcolor[HTML]{d0f0c0} 4.08 & 6.75 \\
         \hline
    \end{tabular}
    \caption{Average algorithm ranking for each combination of input space and design size. Shading indicates the strongest algorithms, and an asterisk indicates common divergence from the target number of points for each full-sub-space causing measurement biases.}
    \label{tab:results}
\end{table}

\begin{figure}[H]
\centering
\includegraphics[width=\linewidth]{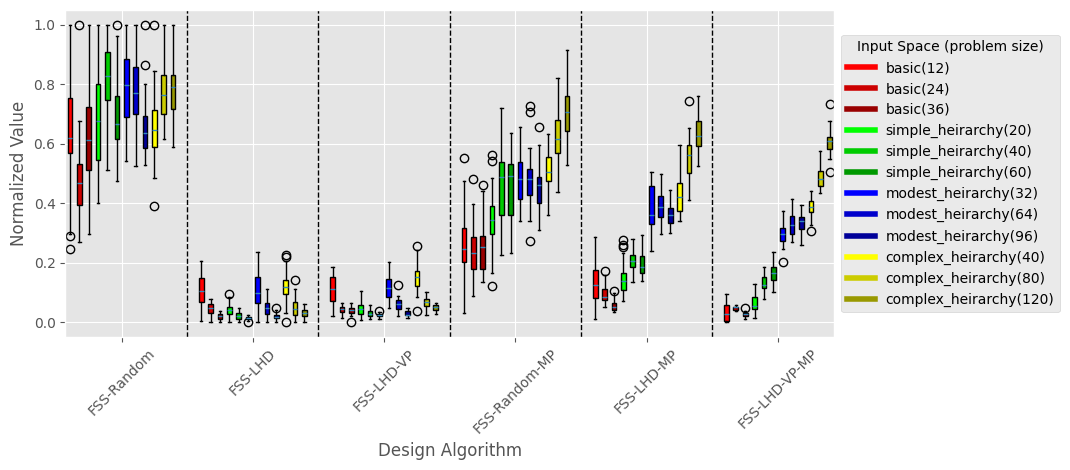}
\caption{\label{fig:weight_discrepancy} Algorithm Space-Filling Results, normalized $M^\text{wdsr}$ (lower is better)} 
\end{figure}

Figure \ref{fig:weight_discrepancy} shows the normalized $M^\text{wdcr}$ for the FSS-based algorithms; the objective is to attain small discrepancies from the uniform distribution from the perspective of every full-sub-space. The left side of the figure shows the algorithm performance, and the right side shows the performance with the MaxPro design optimization. The weighted discrepancies results show that -MP may worsen this criteria. This suggests if subsequent experimentation is expected in a single full-sub-space, then -MP whole-design optimization may not be desired.

\begin{figure}[H]
\centering
\includegraphics[width=\linewidth]{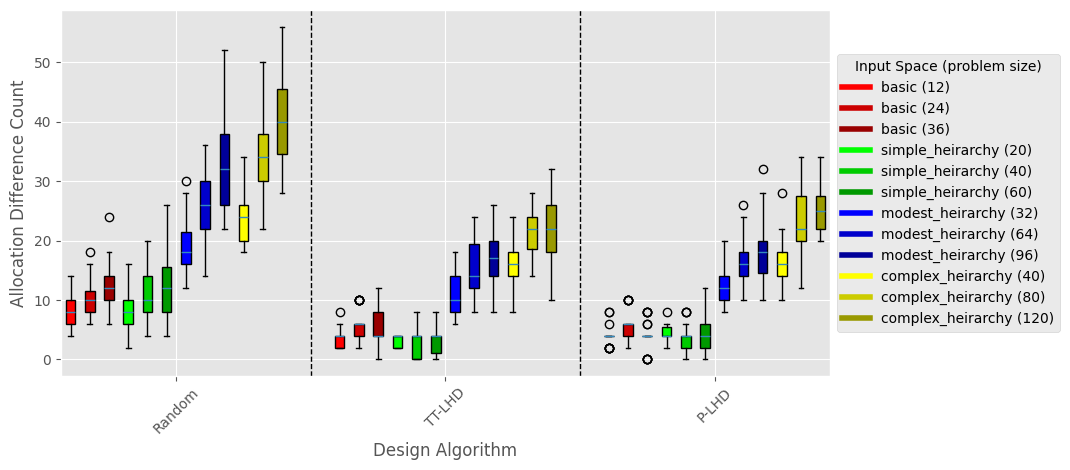}
\caption{\label{fig:allocations_diff} Algorithm Full-Sub-Space Point Allocation Differences from Target Counts (lower is better)}
\end{figure}

Three algorithms (Random, TT-LHD, P-LHD) have an asterisk in table \ref{tab:results}, which marks that they often fail to create designs with point allocations that match the target allocations as derived from the null-portion attributes. The FSS point-allocation to target-allocations differences for these algorithms are shown in figure \ref{fig:allocations_diff}. Random designs inconsistently sample the FSS to match the target point allocations. The FSS-based algorithms, by design, create designs that match the target point allocations.

\begin{figure}[H]
\centering
\includegraphics[width=\linewidth]{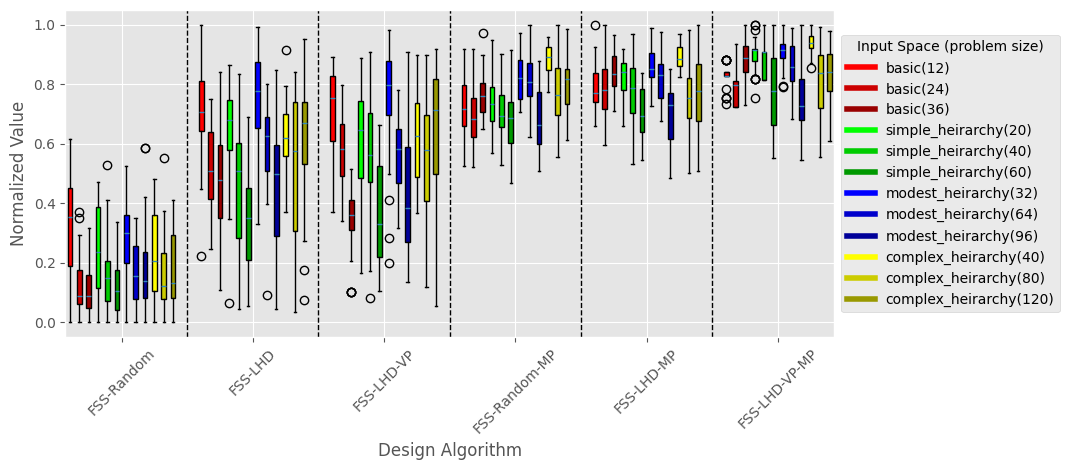}
\caption{\label{fig:min_dist_wo_mp} Algorithm Space-Filling Results, normalized $M^\text{idis}$ (higher is better)}
\end{figure}

Figure \ref{fig:min_dist_wo_mp} shows the FSS-based algorithm designs' normalized minimum interpoint distances over the 30 replications by input-space and design size with-and-without MaxPro processing. With the incorporation of MaxPro processing, designs' $M^\text{idis}$ increases. The FSS-LHP-VP-MP algorithm had the best average for every input spaces and design size. 

\section{Conclusion} \label{sec:summary}

The \proglang{Python} programming language provides a rich set of type hinting and annotation capabilities. \pkg{raxpy} makes it easy to extend these capabilities to design, evaluate, and execute space-filling experiments with optional and hierarchical dimensions. \pkg{raxpy} provides newly proposed design algorithms and measurement techniques, extended to support spaces with optional and hierarchical dimensions. Numeric experimentation highlights differences between the proposed design algorithms and demonstrates the improved space-filling capabilities of the algorithms. 

While \pkg{raxpy} supports creating designs with \code{Union} parameters, future research is suggested to study the space-filling optimizations to support this type of dimension. In a similar manner, \code{List} function parameters provide a similar unexplored possible future research area. Another future research area is the ability to design exploratory experiments for sequential point executions given optional and hierarchical input dimensions. The expressive annotation capabilities of \pkg{raxpy} could be used to simplify optimization experiments of \proglang{Python} functions. We plan to work with the open-source community to refine \pkg{raxpy} given user-desired use cases. For the latest examples, see \url{https://github.com/neil-r/raxpy/tree/main/examples}.

\section{Acknowledgments}
The authors declare no conflicts of interest and confirm that no external funding supported this research. The views and interpretations presented are solely those of the authors and do not necessarily reflect the positions or policies of their affiliated institution(s). We acknowledge and thank Kyle Daley for his support to support code documentation and formatting.

\bibliographystyle{unsrtnat}
\bibliography{ms}  






\end{document}